\documentclass[preprint]{aastex}
\usepackage{natbib}
\usepackage[caption=false]{subfig}
\usepackage{amsmath}

\newcommand{\cmm}{cm$^{-2}$}
\newcommand{\cmmm}{cm$^{-3}$}

\newcommand{\ssfr}{\Sigma_{\rm SFR}}
\newcommand{\msun}{M_\odot}
\newcommand{\mstar}{m_\star}

\newcommand{\tdyn}{t_{\rm dyn}}
\newcommand{\torb}{t_{\rm orb}}
\newcommand{\tcool}{t_{\rm cool}}
\newcommand{\tdep}{t_{\rm dep}}
\newcommand{\tff}{t_{\rm ff}}
\newcommand{\rhostar}{\rho_\star}
\newcommand{\rhostardot}{\dot{\rho}_\star}
\newcommand{\rhogas}{\rho_{\rm gas}}
\newcommand{\sstar}{\Sigma_\star}
\newcommand{\sstardot}{\dot{\Sigma}_\star}
\newcommand{\sgas}{\Sigma_{\rm gas}}
\newcommand{\rhoo}{\rho_0}
\newcommand{\zo}{z_0}

\newcommand{\Ro}{R_0}
\newcommand{\mgas}{M_{\rm gas}}

\newcommand{\mdm}{M_{\rm DM}}
\newcommand{\moo}{M_{200}}
\newcommand{\roo}{r_{200}}

\bibpunct{(}{)}{,}{&}{}{,}

\begin{document}

\title{The Interstellar Medium and Star Formation in Local Galaxies: Variations of the Star Formation Law in Simulations}
\author{Fernando Becerra}
\email{fbecerra@cfa.harvard.edu}
\affil{Departamento de Astronom\'ia, Universidad de Chile, Casilla 36-D, Santiago, Chile}
\affil{Harvard-Smithsonian Center for Astrophysics, Harvard University, 60 Garden Street, Cambridge, MA 02138, USA}
\author{Andr\'es Escala}
\affil{Departamento de Astronom\'ia, Universidad de Chile, Casilla 36-D, Santiago, Chile}

\begin{abstract}
We use the adaptive mesh refinement code Enzo to model the interstellar medium (ISM) in isolated local disk galaxies. The simulation includes a treatment for star formation and stellar feedback. We get a highly supersonic turbulent disk, which is fragmented at multiple scales and characterized by a multi-phase ISM. We show that a Kennicutt-Schmidt relation only holds when averaging over large scales. However, values of star formation rates and gas surface densities lie close in the plot for any averaging size. This suggests an intrinsic relation between stars and gas at cell-size scales, which dominates over the global dynamical evolution. To investigate this effect, we develop a method to simulate the creation of stars based on the density field from the snapshots, without running the code again. We also investigate how the star formation law is affected by the characteristic star formation timescale, the density threshold and the efficiency considered in the recipe. We find that the slope of the law varies from $\sim$1.4 for a free-fall timescale, to $\sim$1.0 for a constant depletion timescale. We further demonstrate that a power-law is recovered just by assuming that the mass of the new stars is a fraction of the mass of the cell $\mstar=\epsilon\rhogas\Delta x^3$, with no other physical criteria required. We show that both efficiency and density threshold do not affect the slope, but the right combination of them can adjust the normalization of the relation, which in turn could explain a possible bi-modality in the law.
\end{abstract}


\section{Introduction}

Galaxy formation is an immensely intricate phenomenon that results from the interaction of multiple processes acting on a wide range of spatial and temporal scales. Among these, one that stands out is star formation, due to its great complexity and the poor understanding. Formation of stars involves a diversity of dynamical, thermal, radiative, and chemical processes and, although it is a local process, it can influence the properties of the galaxy over a broad range of scales. A common model for star formation proposes that gravitational collapse forms dense molecular clouds, out of which stars are created. During this stage, gravity is resisted by many other processes such as rotational shear, turbulence, magnetic fields, and gas pressure. The latter is in turn regulated by the heating processes of the stars. Many theories have been proposed to explain how stars form, including the classic condition for disk instability given by the Toomre $Q$ parameter \citep{Toomre_64, K89, Boissier_03, Heyer_04} and turbulence \citep{Sellwood_99, Kritsuk_02, Wada_02, MacLow_04}. Whatever the correct explanation, all of these theories involve a close relation between star formation and the surrounding interstellar medium (ISM). Several numerical works have been focused on studying this relation \citep{Wada_01, Elmegreen_02, Wada_07,Wang_Abel_09}, but a general consensus has not yet been reached.

Given the complicated interplay between such varied processes, it is astonishing that galaxies follow a regular empirical relation at large scales. A relation between the star formation rate (SFR) and gas density was first proposed by \citet{Schmidt_59} and later extended by \citet{K98} using H$\alpha$, CO, and HI observations of disk galaxies and infrared observations of starburst galaxies. The so-called ``Kennicutt-Schmidt" (KS) relation is a power law of the form $\ssfr \propto \sgas^N$ where $\ssfr$ is the SFR surface density and $\sgas$ the gas surface density. \citet{K98} found that, depending on the scales and tracers used, an index $N$ in the range 1.3-1.6 can be obtained.

Although this model has yielded helpful insights into the comprehension of star formation, recent studies have shown that the formation of stars might be better correlated with molecular gas rather than atomic or total gas surface density \citep{Wong_02, Kennicutt_07, Bigiel_08}. Observations of starburst at high redshift have also postulated that spiral disks at low redshift and starburst galaxies at high redshift follow a KS law with the same slope but different normalizations, in what has been called a ``two sequences" or ``bi-modal" law \citep{Daddi_10, Genzel_10}. Despite the efforts in comprehending this relation and its physical origin, the universality of the star formation efficiency and the slope is still uncertain. Investigating and understanding this relation is thus of critical importance.

Since the seminal work of Kennicutt, numerous works have been done studying this relation, both observation-based \citep{Martin_01, Wong_02, Bouche_07, Kennicutt_07, Leroy_08, Bigiel_08} and simulation-based. Among the numerical works we find simulated disk galaxies in isolation \citep{Li_05, Tasker_Bryan_06, Robertson_08, Gnedin_10}, in mergers \citep{Teyssier_10, Powell_13}, or in cosmological contexts \citep{Springel_03, Kravtsov_03, Governato_04}. In the latter contexts, a common approach is to set up star formation a priori, employing empirical laws such as the KS law, while in the other two techniques are usually implemented: sink particles and sub-resolution recipes. All of them have been successful in reproducing many observational trends, including the KS relation.

\citet{Kravtsov_03} performed hydrodynamic simulations of galaxy formation in a cosmological context. He reproduced a slope of 1.4 for the KS relation using a minimal prescription for star formation. This formula included a constant characteristic gas consumption timescale in such a way that $\rhostardot \propto \rhogas$ and where stellar particles were created with a mass $\mstar=\rhostardot \Delta t_0$ with $\Delta t_0$ representing the global time step. In addition, star formation was allowed to take place only in cells of densities higher than a threshold set to $n_{\rm H}= 50$ \cmmm

Using a sink particle formalism in a smoothed particle hydrodynamics (SPH) simulation of isolated galaxies with isothermal gas, \citet{Li_05} found a slope of 1.5 for the KS relation for two different temperatures. In their simulations sink particles represent star clusters formed in gravitationally bound regions of converging flows that reach densities higher than $10^3$ \cmmm. Mass of sink particles is then converted to star mass assuming a local star formation efficiency of $\epsilon = 50\%$ different from the global efficiency involved in the KS law.

\citet{Tasker_Bryan_06} presented three-dimensional grid-based simulations of local isolated disks. In their work, star formation is defined by three physical criteria following \citet{Cen_Ostriker_92}: a convergent flow, a short cooling time, and Jeans instability, in addition to a density threshold that the cell has to reach in order to create stars. They recovered successfully the slope of the KS law both locally and globally. Their study included simulations of low-density threshold and low efficiency, high-density threshold and high efficiency, with and without feedback.

\citet{Robertson_08} simulated three isolated disk galaxies formed on a cold dark matter (CDM) universe using an SPH code. Their star formation model allow stars to be created in molecular clouds with a mass proportional to the local molecular gas density on a timescale $t_\star$: $\rhostardot \propto f_{\rm H_2} \rhostar/t_\star$. The comparison between SFR and gas surface density shows deviations from the KS relation with indices in the range $n \approx 1.7-4.3$ attributed to radial variation in the molecular gas fraction, the scale height of star forming gas, and the scale height of the total gas distribution. On the other hand, if the comparison is made between SFR and molecular gas surface density, they obtained indices $n \approx 1.2 - 1.5$.

In this paper we present three-dimensional high-resolution hydrodynamic simulations of a local isolated galaxy. The adaptive mesh refinement (AMR) code Enzo allows us to resolve the structure of the multi-phase ISM, recovering many attributes observed in previous numerical works. A similar star formation and feedback treatment from \citet{Tasker_Bryan_06} is implemented with slight modifications. Our aim is to examine star formation and its connection with the surrounding medium. In particular, we carry out a profound analysis of the KS relation, checking its validity at different scales and its dependency on the several physical criteria included in our star formation algorithm. Among them, two particularly significant criteria are star formation efficiency and a density threshold. We then develop a method to calculate an artificial SFR, using the outputs, and we investigate the dependency of the law in a parameter space, varying both quantities, the efficiency and the threshold. The simulation does not include magnetic fields or cosmic-ray pressure, since we want to focus our work on the hydrodynamical evolution of the galaxy.

The paper is organized as follows: We introduce the simulation setup in Section \ref{sec:simulations}, including a description of the code, the initialization of the simulation, and the algorithms used to model star formation and feedback. We then present the analysis of the simulation describing the evolution of the model in Section \ref{sec:evolution}. The analysis of the ISM properties is shown in Section 4 including its structure, probability density functions (PDFs) and powerspectrum. Section 5 presents the study of star formation processes and a detailed description of its relation with the ISM. Finally, we present the summary and conclusions of our work in Section 6.

\section{Simulations}
\label{sec:simulations}

\subsection{The Code}
\label{subsec:code}

We use the hydrodynamic grid-code Enzo\footnote{http://enzo-project.org} \citep{Bryan_14}. Enzo is an Eulerian numerical method based on the structured AMR algorithm by \citet{Berger_Colella_89}. One of the advantages of AMR codes is that they focus the computational effort in regions where it is most useful, allowing different levels of refinement in different regions of the space. Enzo starts covering the simulation box with an uniform grid. Each one of these ``parent" or root grids can also be subdivided into smaller or ``child" grids. The same process can be repeated when a child grid becomes itself a parent grid. The result is a nested structure of grids, where the smaller the grid size, the better the resolution.

The galaxy is simulated in a box of $666 h^{-1}~\text{kpc}$ in comoving coordinates in a $\Lambda$CDM universe with periodic boundary conditions, where we have adopted values $\Omega_m=0.3$, $\Omega_\Lambda=0.7$, and $H_0 = 100h = 67~\text{km}~\text{s}^{-1}~\text{Mpc}^{-1}$ for the matter density, the dark matter density, and the Hubble constant, respectively. The size of the parent grid is $128^3$, and we proceed down to additional seven subgrids of refinement. This level of refinement allows us to reach a resolution of $\sim $40 pc, which is reasonable to resolve a multi-phase ISM in the context of star formation \citep{Ceverino_09}. Two refinement criteria are implemented: refinement by baryon mass if the density of the cell is four times the average density, and refinement by Jeans length to ensure that it is resolved by at least four cells to prevent artificial fragmentation \citep{Truelove_97}.

The hydrodynamic evolution of the gas is calculated using a three-dimensional version of the ZEUS hydrodynamics algorithm \citep{Stone_Norman_92}. Radiative gas cooling is allowed following the curves of \citet{Sarazin_White_87} down to $T=10^4~\text{K}$ and \citet{Rosen_Bregman_95} down to $T=300~\text{K}$. As pointed out by \citet{Ceverino_09}, this range of temperatures will let us resolve the ISM to observe a multi-phase medium.

\subsection{Galaxy}
\label{subsec:galaxy}

We model the galaxy as a three-component system including gas, stars, and dark matter. Gas is discretized using grids, and thus it evolves according to the hydrodynamical equations. On the other hand, stars and dark matter are modeled as external potentials that are fixed in time. Their influence on the gas is therefore only gravitational, changing its velocity and acceleration through the Poisson equation. At the beginning of the simulation we do not add star particles or dark matter particles explicitly, but that does not prevent us from allowing creation of particles once the simulation starts evolving.

We initialize the gas using an exponential profile in the radial direction in cylindrical coordinates and a $({\rm sech})^2$ profile in the vertical direction:

	\begin{equation}
	\rhogas(R,z)=\rho_0 \exp(-R/\Ro) \mathrm{sech}^2\left(\frac{z}{2z_0}\right),
	\label{eq:exponentialprofile}
	\end{equation}

\noindent where $\Ro$ is the disk scale-length, $\zo$ is the disk scale-height and $\rho_0$ is the central volume density. Integrating this expression gives us the relation $\mgas=8\pi \rhoo \zo \Ro^2$ for the total gas mass. We adopt $\Ro=35$ kpc, $\zo=0.4$ kpc and $\mgas=10^{10} \msun$ for our simulation.

The dark matter component is modeled as an external time-independent gravitational field that is fixed through the evolution of the galaxy. We use the popular Navarro-Frenk-White profile \citep[NFW,][]{NFW97} for dark matter density:

	\begin{equation}
	\rho_{\rm DM} (r)=\frac{\rho_{\rm crit} \delta_c}{(r/r_s)(1+r/r_s)^2},
	\label{eq:NFWdensityprofile}
	\end{equation}

\noindent where $r_s=\roo/c$ is a characteristic radius defined as a function of the virial radius $\roo$, $\rho_{\rm crit}=3H_0^2/8\pi G$ is the critical density, $c$ is the concentration parameter, and $\delta_c$ is given by: $\delta_c=200c^3/(3[\ln(1+c)-c/(1+c)])$. The volume integral of Equation \ref{eq:NFWdensityprofile} gives us the dark matter mass profile:

	\begin{equation}
	\mdm (r)=\frac{\moo}{f(c)}\left[ \ln(1+x) - \frac{x}{1+x} \right],
	\label{NFWmassprofile}
	\end{equation}

\noindent where $x=rc/\roo$ and $f(c)=\ln(1+c)-\frac{c}{1+c}$. We adopt a value $c=12$ for the concentration parameter and $\moo=10^{12} \msun$ for the mass enclosed by the virial radius.

Finally, in the case of the star component we use a Miyamoto-Nagai profile \citep{Miyamoto_Nagai_75}, included as an external potential fixed in time, analogous to the dark matter component. This kind of profile resembles an old stellar population present in the form of a disk and a bulge, where the potential and the density are given by \citet{Miyamoto_Nagai_75, Binney_Tremaine_08}:

	\begin{equation}
	\phi(R,z)_{\rm stars} = \frac{-GM_{\rm star}}{\sqrt{R^2+\left( a+\sqrt{z^2+b^2} \right)^2}}
	\label{eq:MNpotential}
	\end{equation}

	\begin{equation}
	\rho_{\rm stars}=\left( \frac{b^2M_{\rm star}}{4\pi} \right) \frac{aR^2+(a+3\sqrt{z^2+b^2})(a+\sqrt{z^2+b^2})^2}{[R^2+(a+\sqrt{z^2+b^2})^2]^{5/2}(z^2+b^2)^{3/2}}
	\label{eq:MNdensity}
	\end{equation}

This potential is an intermediate case between the Plummer potential \citep{Plummer_11} which is recovered in the case when $a=0$, and the Kuzmin potential \citep{Kuzmin_56, Toomre_63} which is retrieved for $b=0$. Depending on the values of $a$ and $b$ the shape of the potential resembles a disk and a bulge. The values selected for our simulation are $a=3.5$ kpc, $b=0.2$ kpc and $M_{\rm star}=4\times10^{10} \msun$

\subsection{Star Formation and Feedback}
\label{subsec:SFandFeedback}

We follow the \citet{Cen_Ostriker_92} algorithm in order to create new star particles. For a cell to enter into this algorithm, its density has to be greater than a density threshold  $\rho_{\rm cell} > \rho_{\rm thres}$. If this requirement is satisfied, then the cell has to fulfill three physical criteria: the gas has to be contracting, the time it takes to cool has to be less that the time it takes to collapse, and it has to be gravitationally unstable. These criteria are represented by Equations \ref{eq:divV}, \ref{eq:tcooldyn}, and \ref{eq:mj} respectively.

	\begin{subequations}
		\begin{equation}
		\vec{\nabla} \cdot \vec{v} <0
		\label{eq:divV}
		\end{equation}
		\begin{equation}
		\tcool < \tdyn \equiv \sqrt{\frac{3\pi}{32G\rho_{\rm tot}}}
		\label{eq:tcooldyn}
		\end{equation}
		\begin{equation}
		m_{\rm cell} > m_{\rm Jeans}
		\label{eq:mj}
		\end{equation}
	\end{subequations}

Once a cell has passed these requirements, a new star is created and its mass is calculated as a function of the star formation efficiency ($\epsilon$), the gas density ($\rhogas$), and the cell volume: $\mstar=\epsilon \rhogas \Delta x^3$. A substantial difference between this expression and the one proposed by \citet{Cen_Ostriker_92} is that we do not consider an efficiency per dynamical time. As a result, there is no delay between the time at which the cell satisfies the conditions and the time when the star particle is created, implying that the cell does not wait a dynamical time $\tdyn$ to turn gas into stars. We adopt this method to prevent the density from continuing to grow during a dynamical time, which would result in stars with unrealistic greater masses. Studies of turbulently regulated star formation predict typical values for the efficiency per dynamical time of the order of a few percent \citep[e.g., ][]{Krumholz_Tan_07}. Although we do not use the same type of star formation efficiency, we stick to this estimation and set our star formation parameters to $\epsilon$ = 3\% and $n_{\rm thres}$ = 0.05 \cmmm. Additionally, a star particle will be created only if its mass is greater than a minimum mass set to $10^5 \msun$. This numerical restriction is introduced to ensure low expenditure of time and memory. In this way we avoid having large amounts of low-mass particles, which would considerably reduce the simulation performance.

We also include stellar feedback to model supernovae explosions. These are treated as an injection of momentum on the surrounding gas. Here we handle feedback as if the star particles were created over a long period of time, so a particle actually loses mass over time in an exponentially decaying way. We use the integral form of \citet{Cen_Ostriker_92} to calculate star mass over time:

	\begin{equation}
	m_{\rm stars}(t)=\mstar \int_{t_{\rm SF}}^t \frac{t-t_{\rm SF}}{\tau^2}\exp\left[\frac{-(t-t_{\rm SF})}{\tau}\right]dt.
	\end{equation}

An energy equivalent to a supernova of $10^{51}$ ergs is then injected for every 55 $\msun$ of stellar mass formed. We do this through the increase of the momentum of the surrounding material. This kind of feedback has been postulated as one of the most important processes in self-regulation of star formation, decreasing considerably the amount of stars formed during the evolution of the galaxy \citep{Tasker_Bryan_06, Hummels_12, Hopkins_13}.

\section{Evolution of the disk}
\label{sec:evolution}

	\begin{figure}[h!]
	\begin{center}
	\includegraphics[scale=0.38]{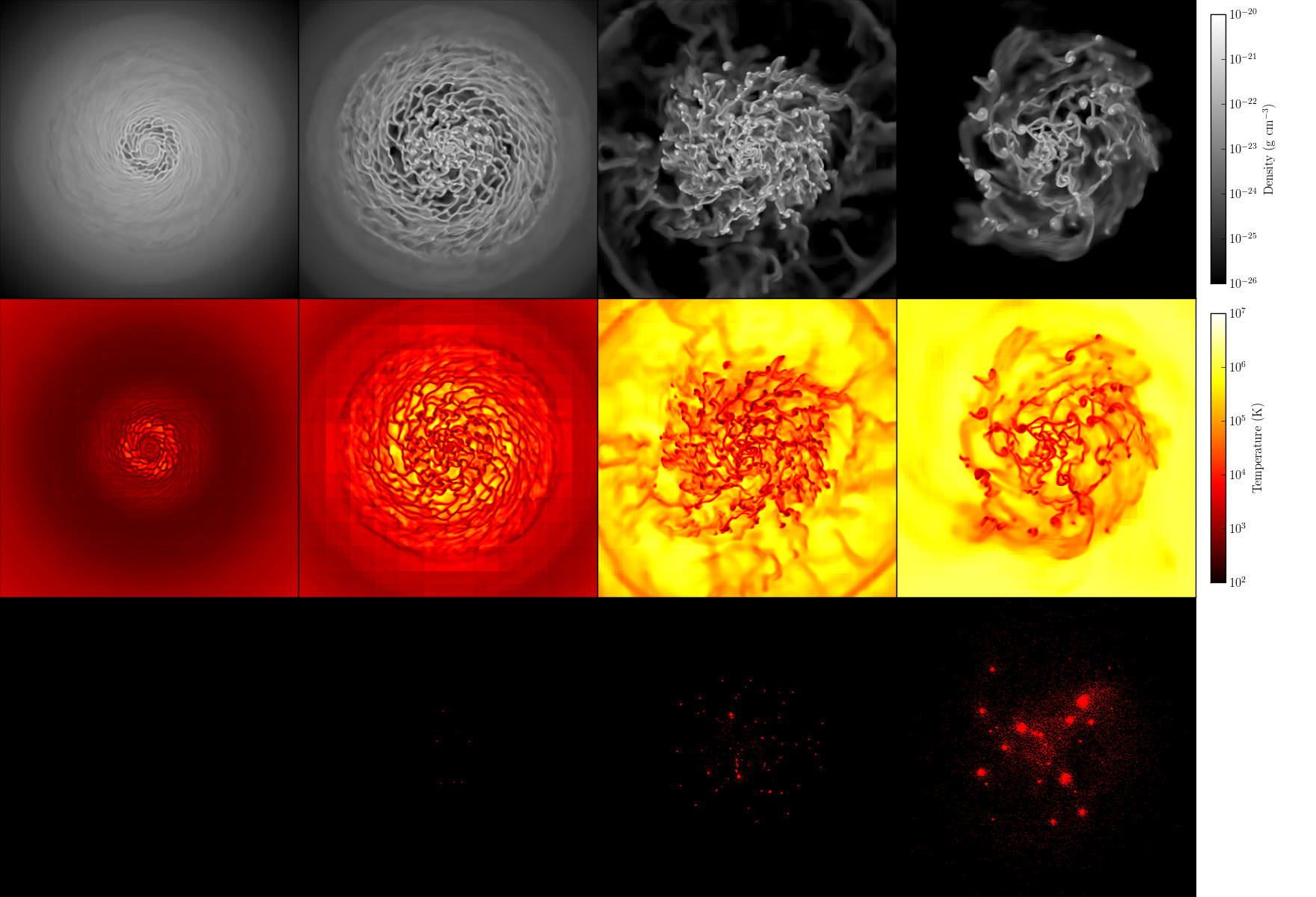}
	\caption{Time evolution of our model. The upper row shows density projections, the middle row displays temperature projections, and the position of stars are shown in the lower row. From left to right, columns represent snapshots of 30 kpc wide at times $t$ $\sim$100, 200, 400, and 800 Myr. Fragmentation starts at the center of the disk to soon spread outward. The final state after $\sim$1 Gyr is a fragmented and highly turbulent disk.}
	\label{fig:evolution}
	\end{center}
	\end{figure}
	
We let the simulation evolve for a total time of $\sim$ 1 Gyr. In Figure \ref{fig:evolution} we present the evolution of the disk in a series of face-on snapshots. Columns from left to right correspond to times $t$ $\sim$ 100, 200, 400, and 800 Myr. The upper row shows density projections in a scale from $1 \times 10^{-26}$ g \cmmm to $1\times10^{-20}$ g \cmmm, the middle row exhibits temperature projections in a scale from $10^2$ K to $10^7$ K, and the lower row displays the position of particles (stars) created so far. Due to the shorter dynamical time, fragmentation starts in the densest regions located in the central part of the disk and soon spreads outward with time. This behavior is expected, since the Toomre $Q$ parameter \citep{Toomre_64} is lower than unity at the beginning of the run, as shown later in Section \ref{subsec:stability}. At the same time, the disk departs from the initial state of constant temperature, allowing denser structures to cool and form a high-temperature intra-filament medium. As a result of fragmentation, the disk is characterized at final stages by a complex mixture of high-density and low-temperature filaments and clumps surrounded by warm and hot gas. Star formation begins at around $t$ $\sim$ 200 Myr in high-density clumps, once the gas density has reached the density threshold and the cells have fulfilled the physical conditions described in Section \ref{subsec:SFandFeedback}. From then on, supernova explosions play an important role in heating the surrounding medium and producing low-density zones, while the shocks generated by the same process trigger star formation in high-density regions. By the end of the run the simulation has reached a quasi-stationary state, where a fragmented and highly turbulent disk is observed, consistent with previous simulations \citep{Wada_01, Tasker_Bryan_06}. In that time the gas has done more than four orbits around the center at a radius enclosing 95\% of the stars ($R_{95}$) and roughly 25\% of the gas has been consumed. Additionally, the total particle mass does not get higher than $7 \times 10^9 \msun$, thus the stellar external potential dominates during the whole evolution.

\subsection{Stability}
\label{subsec:stability}

A good estimation of the dynamical stability of the simulated disk is characterized by the Toomre $Q$ parameter \citep{Toomre_64, Goldreich_65} given by \citep{Leroy_08}:

	\begin{equation}
	Q=\frac{\kappa \sigma}{\pi G \sgas},
	\label{eq:toomreQ}
	\end{equation}

\noindent where $\kappa$ is the epicyclic frequency, $\sgas$ the gas surface density, and $\sigma$ the gas velocity dispersion. The Toomre $Q$ parameter gives a criteria to determine whether the galaxy will globally fragment and therefore create bounded collapsing clumps. In disk galaxies, a critical average value of $Q_{\rm crit} \sim 1$ has been found for a two-dimensional disk \citep{Toomre_64} and a slightly lower value $Q_{\rm crit} \sim 0.7$ for a finite disk thickness \citep{Goldreich_65}. If $Q$ is greater than this critical value, the disk will be stable; on the contrary, if it is less, then it will be unstable and hence susceptible to fragmentation.

	\begin{figure}[h!]
	\begin{center}
	\includegraphics[scale=0.5]{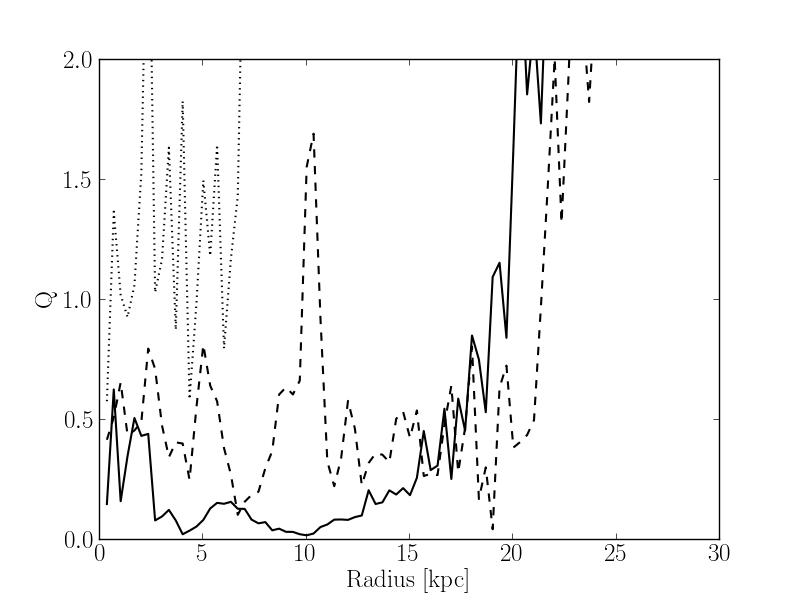}
	\caption{Evolution of the Toomre $Q$ parameter for our galaxy model. Solid, dashed, and dotted lines are $Q$ profiles at $t$ $\sim$ 100, 200, and 400 Myr, respectively. As the simulation evolves, the median value of $Q $increases and the radius at which $Q$ surpasses 2 decreases. This indicates that the disk becomes more stable, especially in outer regions.}
	\label{fig:Qvst}
	\end{center}
	\end{figure}

The evolution of the Toomre parameter for our model is shown in Figure \ref{fig:Qvst}. The lines represent the profile of $Q$ at $t \sim$100 Myr (solid line), 200 Myr (dashed line), and 400 Myr (dotted line). Initially, our disk have $Q < 1$ for radius less than 20 kpc, indicating that this region is unstable and therefore will fragment. This prediction can be compared with the first and second columns in Figure \ref{fig:evolution}, where it is evident that fragmentation and formation of clumps occur where the Toomre parameter is lower than unity. Additionally, this guess is reinforced by snapshots at later stages of the simulation, from which we deduce that a radius of 20 kpc encloses at least 95\% of the stars created during the evolution of the galaxy.

From Figure \ref{fig:Qvst} we can distinguish two clear tendencies: (1) the average value of the parameter increases as time goes by, beginning at values $Q \ll 1$ and increasing to 0.1 at $t = 100$ Myr, 0.5 at $t = 200$ Myr, and reaching 1.2 at $t$ = 400 Myr; and (2) as the disk evolves the radius beyond which star formation does not occur decreases. We extract from the plot that its value (defined as the size at which the Toomre parameter escalates to values greater than 2) decreases from $\sim$23 kpc at the start to $\sim$7 kpc at the end of the run. This last feature is easy to see from Figure \ref{fig:evolution}, where gas has become smoother and more stable in the outer regions of the simulation, and hence star formation has been suppressed there. This is expected since the disk has reached a steady state at later times of the run, and our results are in agreement with previous studies \citep{Wada_99, Wada_01}.

\section{Interstellar Medium}
\label{sec:ISM}

\subsection{Structure}
\label{subsec:structure}

\begin{figure}[h!]
	\begin{center}
	\includegraphics[scale=0.5]{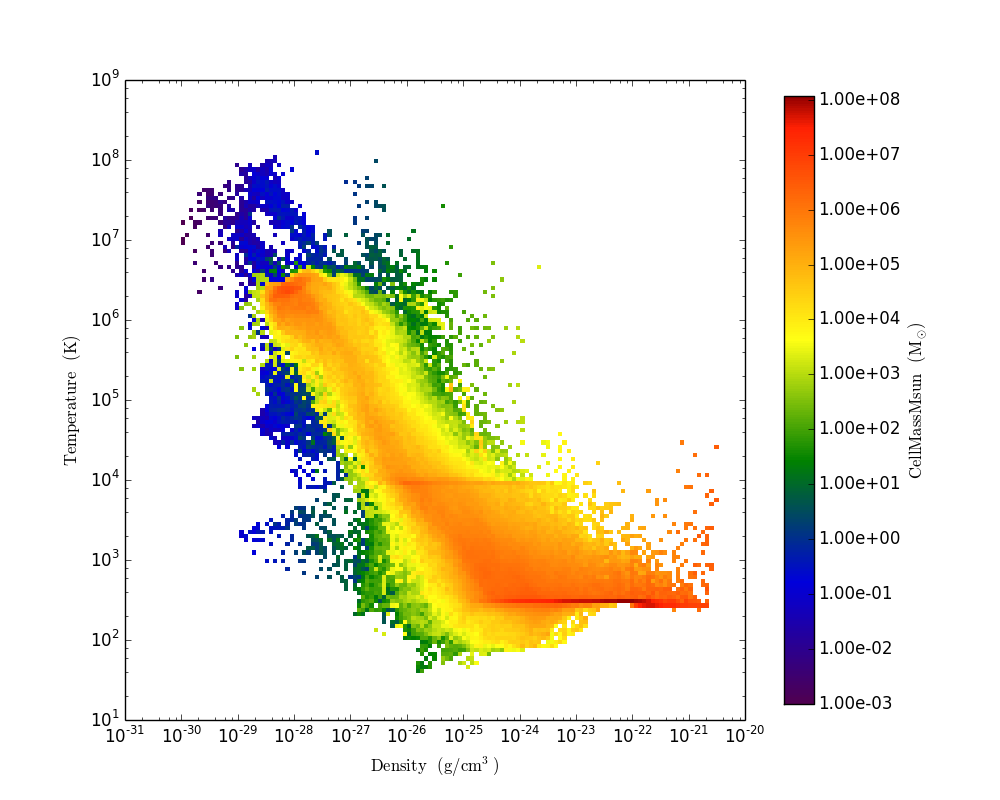}
	\caption{Gas volume density and temperature phase diagram for our model, weighted by cell mass. Three phases are easily distinguishable: a high-density cold phase (bottom right), a medium-density warm phase (center), and a low-density hot phase (top left).}
	\label{fig:phase}
	\end{center}
	\end{figure}

In the late 1970s, \citet{McKee_Ostriker_77} postulated a three-component ISM that includes the effects of supernova explosions. Recent studies have extended this concept to a wide range of phases, but the idea of a multi-phase ISM has remained intact. From the mass-weighted phase space presented in Figure \ref{fig:phase}, it can be seen that our simulation recovers successfully a three-phase ISM defined by the red regions in the plot. First, a high-density cold gas with gas surface densities in the range from $10^{-3}$ to $10^{-1}$ g \cmm~and $T \sim 300$ K is seen at the bottom right of the phase space. We associate this phase with the collapsed filaments and clumps which are easily distinguishable in Figure \ref{fig:evolution}. Their high densities enhance the radiative cooling at which they are exposed, provoking a drop in their temperatures and forming a cold gas phase as a consequence. A second phase is dominated by the warm gas, with temperatures close to $10^4$ K and densities between $10^{-5}$ and $10^{-4}$ g \cmm~located close to the center of the plot. We link this warm phase to the gas in the vicinity of cold filaments and clumps. Their densities are not high enough to collapse and form denser structures, so only a moderate radiative cooling acts on them. Finally, a third phase of low-density hot gas can be identified in the region situated in the upper left. We retrieve a characteristic temperature of $T \sim 10^6$ K and a characteristic gas surface density of $\sgas \sim 10^{-6}$ g \cmm. In the central regions, this is a product of supernovae explosions that heats the surrounding gas, increasing its temperature and producing low-density blobs between fragments. The gas in the outer regions has remained stable during the whole evolution and thus has not formed high-density structures. Simultaneously, feedback from supernovae reach these regions, heating the gas up and contributing to the third phase of the ISM.

\subsection{Probability Density Functions}
\label{subsec:PDFs}

\begin{figure}[h]
		\centering
		\makebox[\textwidth]{
		\subfloat[Density PDF]{
		\label{fig:densitypdf}
		\includegraphics[width=0.54\textwidth]{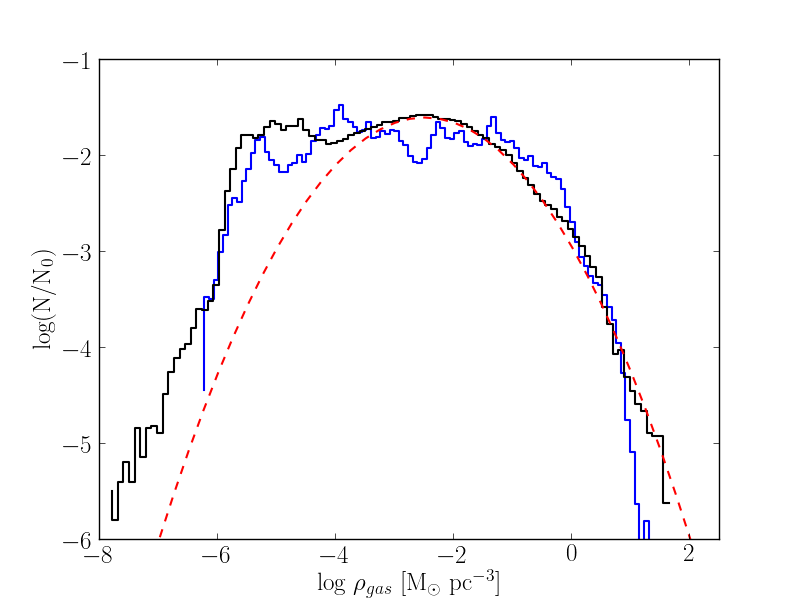}
		}
		\subfloat[Mass PDF]{		
		\includegraphics[width=0.54\textwidth]{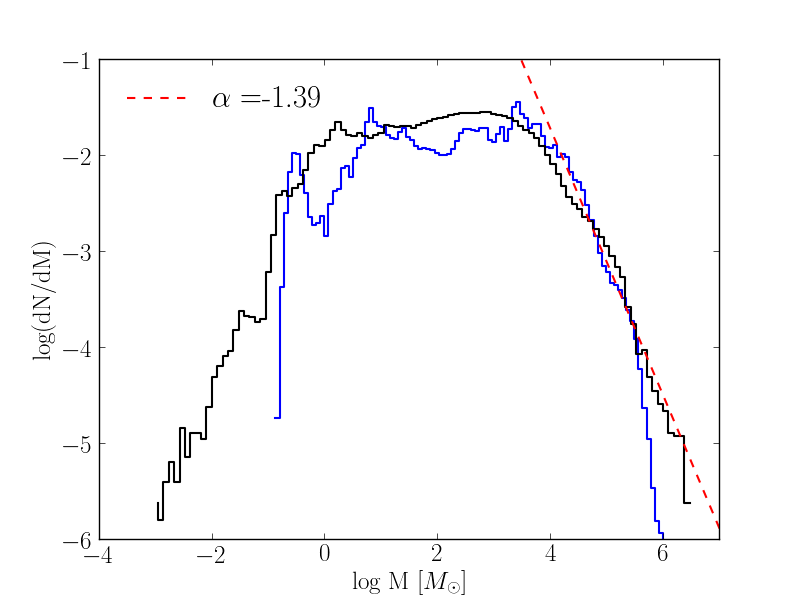}
		\label{fig:masspdf}}
		}
		\caption{Density (left) and mass PDFs (right) for our simulation. The solid blue line shows PDFs at $t \sim 100$ Myr, while the black solid line represents $t \sim 1$ Gyr. Red dashed lines show our best lognormal fit for the high-density range in the left, and the best power law fit to the mass PDF in the high-mass regime in the right. In the case of the density PDF, the lognormal fit is characterized by $\rhoo = 10^{-2.5}\msun \text{pc}^{-3}$ and $\sigma_0 = 10^{1.0}$, while in the mass PDF case a power law of index $\alpha = -1.39$ is obtained (see text for details).}\label{fig:pdfs}
	\end{figure}

Figure \ref{fig:pdfs} shows the gas density (left) and the mass PDF (right) after 100 Myr (blue solid line) and after 1 Gyr (black solid line). During the initial phase, the disk has just started to fragment, and hence the density PDF (Figure \ref{fig:densitypdf}) is highly influenced by the initial exponential profile showing a narrow and flat distribution. Later in the evolution, fragmentation starts playing a substantial role, provoking the gravitational collapse of structures and forming high-density cells. As stars form in those cells, explosions from supernova feedback create low-density regions. The outcome of both processes is a wider density distribution. The high-density tail of the PDF can be fitted by a lognormal PDF (LN-PDF) over almost four orders of magnitude in gas density (red dashed line) in concordance with simulations of \citet{Wada_01, Kravtsov_03, Wada_07, Tasker_08}. The lognormal fit in gas density is given by the expression:

	\begin{equation}
	f(\rho)d\rho=\frac{1}{\sqrt{2\pi\sigma_0^2}}\exp{\left[ -\frac{\ln{(\rho/\rhoo)}^2}{2\sigma_0^2} \right]}d\ln{\rho}
	\end{equation}

\noindent where $\rhoo$ is the characteristic volume density scale and $\sigma_0^2=\ln{(1+b\mathcal{M}^2)}$ is the dispersion of the LN-PDF, with $\mathcal{M}$ being the Mach number. The average density in our best-fit LN-PDF is $10^{-2.5} \msun \text{pc}^{-3}$, while the dispersion is close to $10^{1.0}$.

\citet{Wada_07} argued that this distribution is a result of the non-linear interaction of many random and independent processes happening in the ISM, for example: mergers, collisions and tidal interactions between clumps/filaments, compression by shock/sound waves, galactic shear, and turbulence. The randomness of these events allows us to express the density as the result of the multiplication of those factors, which in logarithmic space corresponds to a sum. Considering infinite random processes, a Gaussian naturally emerges according to the central limit theorem. Other authors have suggested that an LN-PDF for gas density is a natural outcome of isothermal gas \citep[e.g.,][]{Passot_98, Vazquez_Semadeni_00}. Although the densest cells in our model have temperatures near 300 K, our simulation is characterized by a multi-phase medium with temperatures generally lying in a range from $10^2$ K to $10^4$ K in the high-density range (Figure \ref{fig:phase}). We thus obtain a lognormal fit at high densities, even though our simulation is not isothermal. It is believed that this LN-PDF could be the origin of the KS relation \citep{Elmegreen_02, Wada_07} which will be studied in detail in Section \ref{sec:SF}.

A curious feature is the peak in the low-density part of the histogram shown in Figure \ref{fig:densitypdf}. We attribute this to supernovae explosions that generate a low-density hot phase in the ISM, reemphasizing the key role that they play in the self-regulation of the ISM. A similar attribute was also obtained by \citet{Slyz_05}, although they found a bi-modal PDF for simulations that includes feedback, which is not seen in our case.

	\begin{figure}[h!]
	\begin{center}
	\includegraphics[scale=0.5]{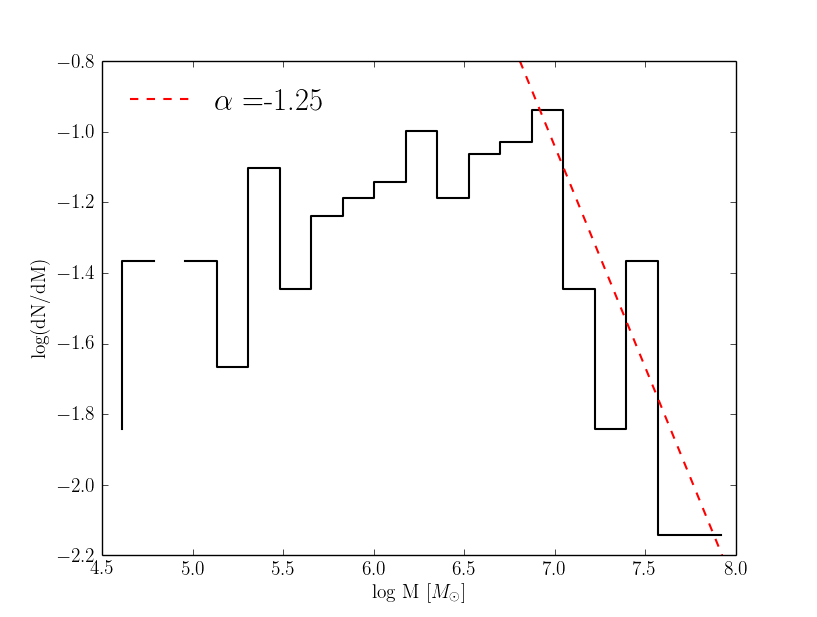}
	\caption{Mass histogram for clumps at the end of the simulation. The red dashed line represents the best fit to the high mass range ($\log{M_{\rm clumps}} > 7.0$), characterized by a slope of $\alpha = -1.25 \pm 0.39$.}
	\label{fig:clumpspdf}
	\end{center}
	\end{figure}

Understanding the mass distribution of stars and molecular clouds is another challenge in the study of the ISM. Unfortunately, given the restriction of a mass threshold imposed in our star formation recipe, we are not able to get a broad sample of stellar masses, and hence a study of star mass distribution is beyond our scope. However, in our model we are still able to study the mass distribution of cells and clumps. We proceed to show the mass PDF of cells for times $t \sim 100$ Myr (blue solid line) and $t \sim 1$ Gyr (black solid line) in Figure \ref{fig:masspdf}. Because of our refinement criteria, cells with higher masses will also be smaller in size and hence have greater densities, while larger cells will be less massive and less dense. We then focus on the high-density range of the distribution. Unlike \citet{Ballesteros_Paredes_06}, we do get a distribution well fitted by a power law at densities greater than 10 \cmmm, obtaining $dN \propto M^{-2.15}dM$ at $t \sim 100$ Myr and $dN \propto M^{-1.39}dM$ at $t \sim 1$ Gyr. The change of slope can be understood from gravitational collapse which allow cells with masses $\sim 10^4\msun$ to form more massive cells. As a result, low-mass cells move to the high-mass regime in the PDF, causing a decrease in the number of cells with $m_{\rm cell} = 10^4\msun$ and an increase of cells with masses $\sim 10^6\msun$.

Studying the mass PDF at higher scales involves using a clump finder algorithm to locate independent objects in space and velocity. The resulting clumps have characteristic masses of $M_{\rm clumps} \sim 10^{4.5-8.0} \msun$, and densities between 1 and 2000 \cmmm. We then proceed to plot the mass histogram for these clumps in Figure \ref{fig:clumpspdf}. We recover a high-scatter power-law consistent with a slope $\alpha = -1.25 \pm 0.39$ for the high clump mass range ($\log{M_{\rm clumps}} > 7.0$), where clumps have masses comparable to giant molecular associations. For reference, our result is slightly smaller than the slope of 1.5 for clouds mass function in the range $M \gtrsim 10^5 \msun$ found in previous studies \citep{Williams_97, Rosolowsky_05, Blitz_07, Padoan_07}. A power law PDF is thought to be the outcome when the nonlinear advection operator term dominates in the hydrodynamic equations \citep{Scalo_98a}.

\subsection{Power Spectrum}
\label{subsec:ps}

\begin{figure}[h!]
	\begin{center}
	\includegraphics[scale=0.5]{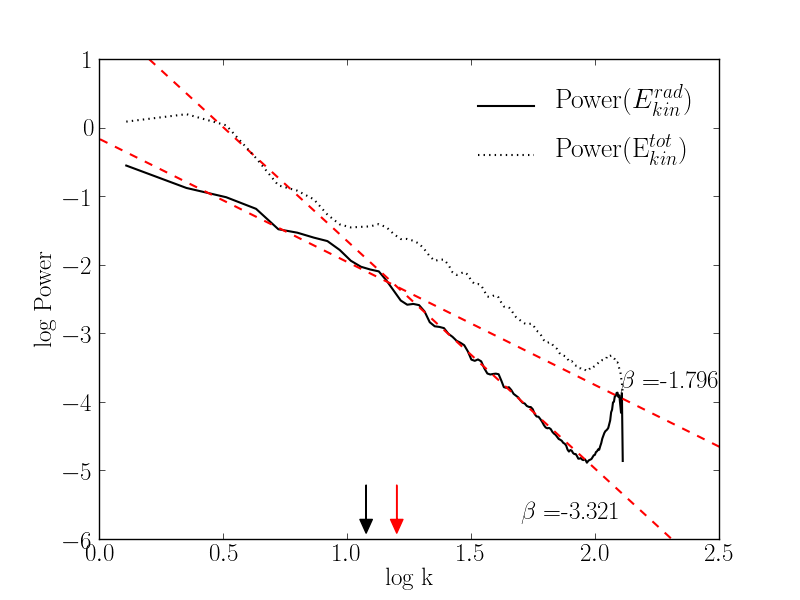}
	\caption{Kinetic energy power spectrum $E(k)$ for our simulation at the end of the run. Black solid line is radial kinetic energy power spectrum, while black dotted line is total kinetic energy power spectrum. Red dashed lines are fits for small and large scales. The red arrow at the bottom indicates the wave number at which both fits intersects. The black arrow points out the intersection of a Kolmogorov regime for large scales and a regime of the type $E(k) \sim k^{-3}$ for small scales (see text for details).}
	\label{fig:ps}
	\end{center}
	\end{figure}

The power spectrum of the simulation at time $t \sim 1$ Gyr is displayed in Figure \ref{fig:ps}. Total kinetic energy spectrum is plotted in black dashed line, and black solid line is used for radial kinetic energy power spectrum. Both profiles have a similar slope, being almost parallel in the whole wave number range, presenting in average a shift in the $y$-axis of $\sim$0.5. This displacement comes from the assumption of an isotropic turbulent velocity. In such a case the relation $\langle v^2_r\rangle = \langle v^2_\theta\rangle = \langle v^2_z\rangle$ is valid, and then we can calculate the average velocity square as $\langle v^2\rangle = \langle v^2_r\rangle + \langle v^2_\theta\rangle + \langle v^2_z\rangle = 3\langle v^2_r\rangle$. Finally, we can deduce that $\log{\langle v^2\rangle} \approx \log{3} + \log{\langle v^2_r\rangle}$, where $\log{3}$ can be approximated as $\sim$0.5.
	
In Figure \ref{fig:ps} the two red dashed lines represent a fit to the power spectrum at small and large scales. The red arrow at the bottom indicates the position in the $x$-axis where both fits intersect, showing that at approximately $1.2\Ro$ (the disk scale height, see Equation \ref{eq:exponentialprofile}) the power spectrum changes its slope. If we assume instead that at large scales the power spectrum is well represented by Kolmogorov turbulence with a power law $E(k) \sim k^{-5/3}$ \citep{Kolmogorov_41}, while at small scales it is best fitted by a law of the type $E(k) \sim k^{-3}$, we obtain that the intersection of both fits is now $\sim 1.6\Ro$, indicated by a black arrow to the left of the red arrow. This scheme is known as a double cascade, which was first proposed by \citet{Kraichnan_67} in two-dimensional turbulence and recently discussed by \citet{Bournaud_10} in the context of a LMC-like galaxy simulation. In such a case, energy is injected at scales $\sim R_{\rm disk}$, provoking an inverse energy cascade to larger scales and a direct enstrophy cascade to smaller scales, where enstrophy is defined as the integral of the square of the vorticity $\epsilon=\int_S ||\vec{\nabla} \times \vec{v}||^2$. Assuming that interactions act to produce equilibrium, enstrophy is transferred to high wavenumbers where it is dissipated by viscosity (direct enstrophy cascade). The opposite happens with energy, which is transported to low wavenumbers (inverse energy cascade). At scales smaller than the disk scale height, the problem becomes three dimensional and the double cascade theory is no longer valid. At those scales, supernovae explosions inject energy on the gas, producing a peak in the powerspectrum, although this can not be concluded for sure because of the lack of resolution in our simulation.

\section{Star Formation}
\label{sec:SF}

One of the first pioneering works studying star formation and developing a star formation law was proposed by \citet{Schmidt_59, Schmidt_63}. He suggested that the relation between SFR volume density and gas volume density was a power law of index $n$ between 1 and 2. Almost four decades afterwards, \citet{K98} derived the exponent and the normalization of the relation using H$\alpha$, CO, and HI observations of disk galaxies and infrared observations of starburst galaxies. The so-called KS law for star formation is then given by:

	\begin{equation}
	\ssfr = (2.5 \pm 0.7) \times 10^{-4} \times \left( \frac{\sgas}{1\msun \text{pc}^{-2}} \right)^{1.4 \pm 0.15} \msun \text{yr}^{-1} \text{kpc}^{-2}
	\label{eq:kslaw}
	\end{equation} 

In this section, we study star formation history and the behavior of the KS relation in our model galaxy.

\subsection{Star Formation Rate}
\label{subsec:SFR}

\begin{figure}[h]
	\begin{center}
	\includegraphics[scale=0.5]{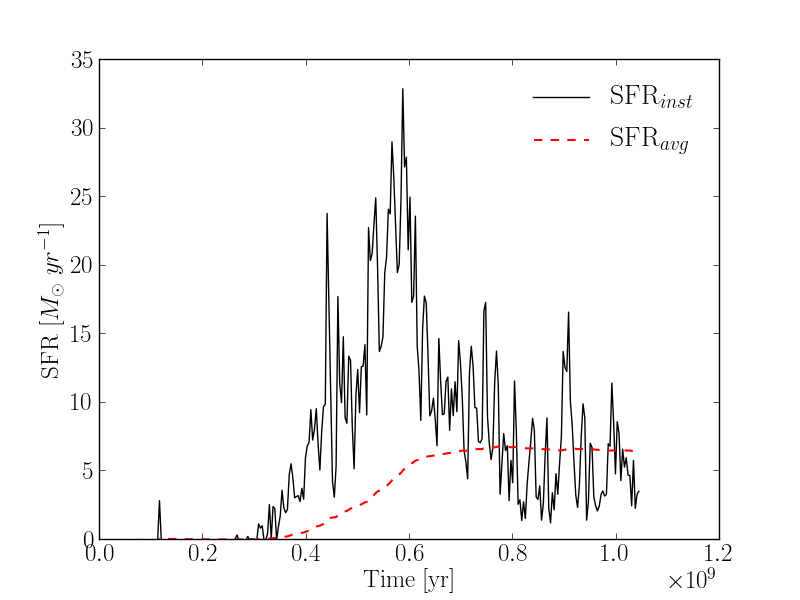}
	\caption{Star formation rate as a function of time. The instantaneous SFR curve is shown in black solid lines, while red dashed lines show the average SFR (see text for definitions of both quantities). A prominent peak at $t \sim 6 \times 10^8$ yr is distinguished in the case of SFR$_{\rm inst}$. During that time, SFR$_{\rm avg}$ experiences a sustained rise. From then on, both values reach a quasi-stationary state.}
	\label{fig:sfr}
	\end{center}
	\end{figure}

Figure \ref{fig:sfr} represents the SFR following two approaches: (1) using the instantaneous SFR (SFR$_{\rm inst}$) defined as the star mass created between two successive timesteps (black solid line), and (2) using the average SFR (SFR$_{\rm avg}$) defined as the cumulative star mass created from the beginning of the simulation (red dashed line) divided by the total time. Both quantities are useful when compared to observations but, depending on the nature of the observational data, one might be more appropriate than the other. For instance, if SFRs are independent of any feedback processes given the density distribution of the ISM, then the SFR$_{\rm inst}$ fits better. On the other hand, SFR$_{\rm avg}$ might be more suitable when SFR is derived from integrated quantities.

A highly variable instantaneous SFR that reaches a peak around $t \sim 550 - 600$ Myr and then decreases slowly to a stationary value is described by the black line. Local peaks are also observed and correspond to explosive periods of star formation driven by supernova feedback. The red dashed line shows an average SFR with a continuous growth in the first half of the run, which reaches a constant value of SFR$_{\rm avg} \sim 6~\msun \text{yr}^{-1}$ around $t \sim 700$ Myr. Prominent differences can be noted between 400 Myr and 600 Myr, when stellar feedback starts to become important and generates additional bursts of star formation. After this stage both, instantaneous and average SFR, arrive at a roughly constant value, and differences are due to transient star formation periods. It is worth noting that, despite the different time ranges involved in the definitions, they reach comparable values in the quasi-stationary state. At times greater than 1 Gyr, it is expected that both SFR$_{\rm inst}$ and SFR$_{\rm avg}$ will tend to decrease as long as gas is converted into stars, eventually reaching zero once the gas has been completely consumed.

\subsection{Star Formation Laws}
\label{subsec:SFL}

\begin{figure}[h]
	\begin{center}
	\includegraphics[scale=0.47]{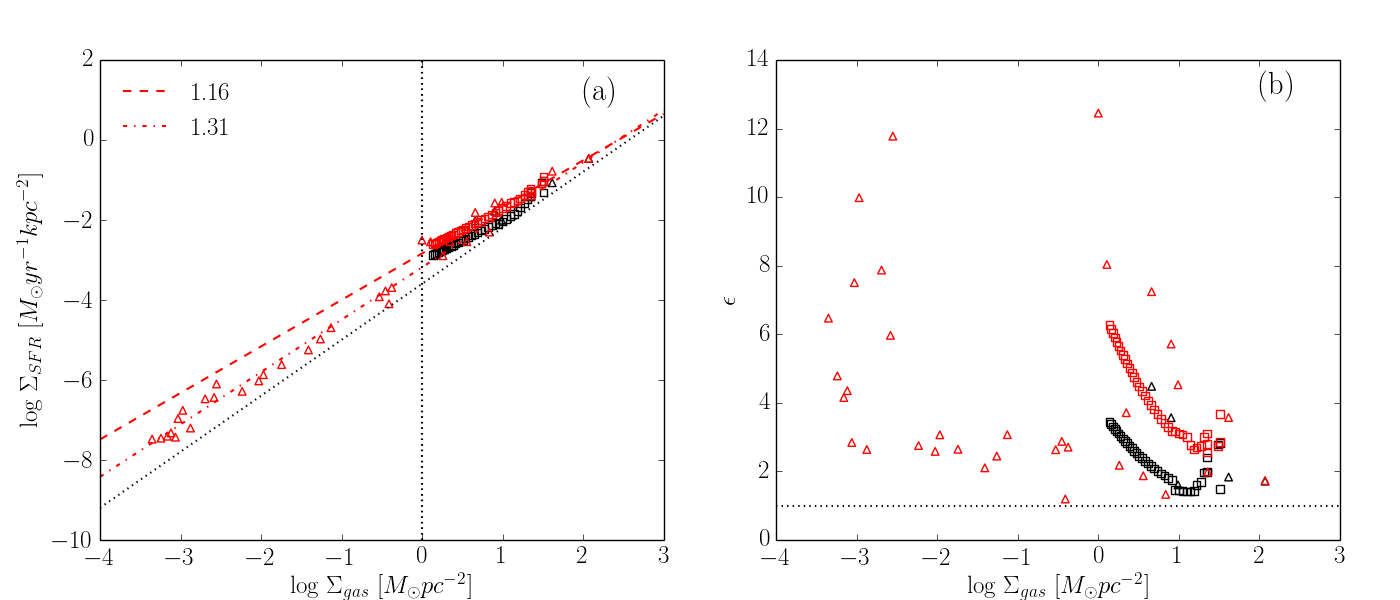}
	\caption{Kennicutt-Schmidt law at $t \sim 1$ Gyr. Panel (a) shows SFR surface density versus gas surface density calculated for disks (squares) and rings (triangles). Colors represent different approaches to compute $\ssfr$, plotting the instantaneous SFR in black and the average SFR in red. The original \citet{K98} relation is also included in black dotted lines. The vertical dotted line indicates the density at which a break in the power law is observed. The best fit to the points above and below that value are shown in red dashed and dashed dotted lines respectively. Panel (b) shows the ratio of the calculated $\ssfr$ to the value predicted by the \citet{K98} original fit (see text for details).}
	\label{fig:kslaw}
	\end{center}
	\end{figure}

In Figure \ref{fig:kslaw} we study how the KS relation behaves at $t \sim 1$ Gyr using the instantaneous (in black) and the average SFR (in red). Following previous works in the field, we have calculated the SFR using disks (squares) and rings (triangles), with their origins located at the center of the galaxy (see, for example, \citet{Tasker_Bryan_06}). In panel (a) we plot the SFR surface density versus gas surface density, and the original KS fit in black dotted lines. To better appreciate the difference between both values, panel (b) shows the ratio, $\epsilon$, of the $\ssfr$ calculated from the outputs to the predicted value of the KS fit.

Differences between average and instantaneous SFRs are better appreciated when we examine disks. By keeping the gas surface density invariable, we can distinguish an overall trend $\sgas^{\rm inst} < \sgas^{\rm avg}$ for all points. In Figure \ref{fig:sfr} we see that during late stages of the evolution both values have reached a quasi-stationary value, and that the instantaneous SFR oscillates around the average SFR. Although there is discrepancy between red and black squares, there is no significant difference in the slope of the relation at those scales. If we compare two disks with distinct radius, the one with the higher radius encloses more star and gas mass. However, at higher radius the gas becomes less dense and less massive, implying that the mass of gas added is less than the mass of stars added. Therefore, increasing the radius of disks will be reflected in a slower decrease in SFR than gas surface density, resulting in the linear trend seen in the plot.

Although in general all points lie close to the KS fit, the plot suggests that they follow power-law relations with different exponent for low and high gas surface densities. This break in the relation has been proposed by observational works in the last decade \citep{Bigiel_08}, and it is believed to be caused by the transition from atomic to molecular hydrogen at $\sgas = 1~\msun \text{pc}^{-2}$ (vertical dotted line). We then proceed to derived the best fit to the points with gas surface density smaller than $1~\msun \text{pc}^{-2}$, and with $\sgas$ greater than that value, showed in red dashed dotted and dashed lines respectively. The slopes of the fits are included at the upper left corner of the diagram. 

For the high gas surface density regime, our slope of 1.16 is close to the value observed by \citet{Bigiel_08} ($n \sim 1.0$) and comparatively smaller than the original KS law. In the other regime, we obtained a slope of $n = 1.31$, which is more consistent with the original KS relation, but smaller than the values derived by \citet{Bigiel_08} for the same range. Despite the fact that we have not considered a distinction between atomic and molecular hydrogen in our simulation, we do see a break in the power law in our plot, suggesting that the differences in the exponent are not only due to the configuration of the hydrogen, but to a more fundamental property, such as the size of the regions being considered. 

\begin{figure}[p]
	\begin{center}
	\includegraphics[scale=0.8]{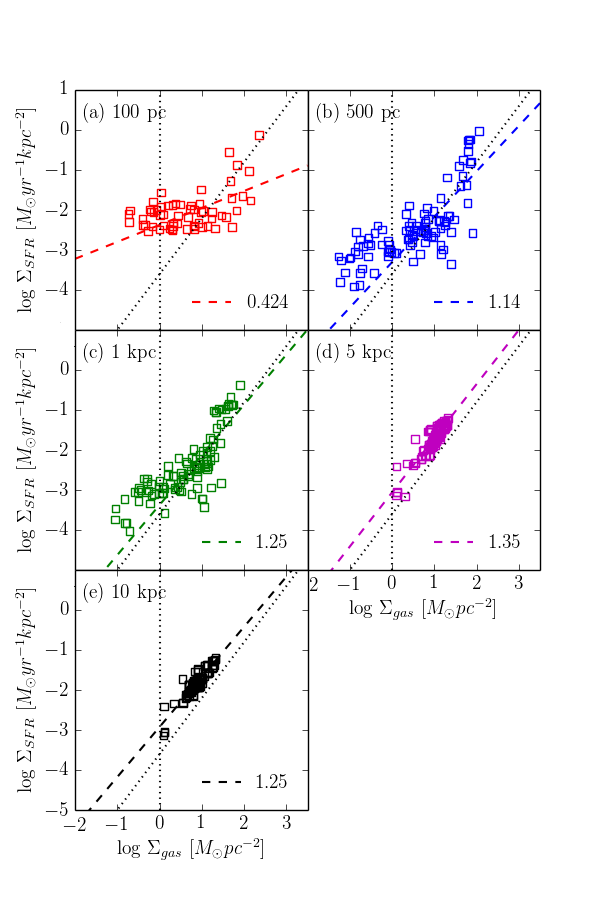}
	\caption{Kennicutt-Schmidt relation using different region sizes. Squares represent values for disks random located, with a radius indicated in each panel. Vertical dotted line indicates $\sgas = 1 ~\msun \text{pc}^{-2}$, and the inclined dotted line shows the original \citet{K98} fit. Dashed lines are best fit lines to the points, with their slopes showed at the lower right corner of each panel. From the plot it is clear that as the radius increases, the slope of the relation tends to $\sim$1.3.}
	\label{fig:kssizes}
	\end{center}
	\end{figure}

\begin{figure}[h]
	\begin{center}
	\includegraphics[scale=0.4]{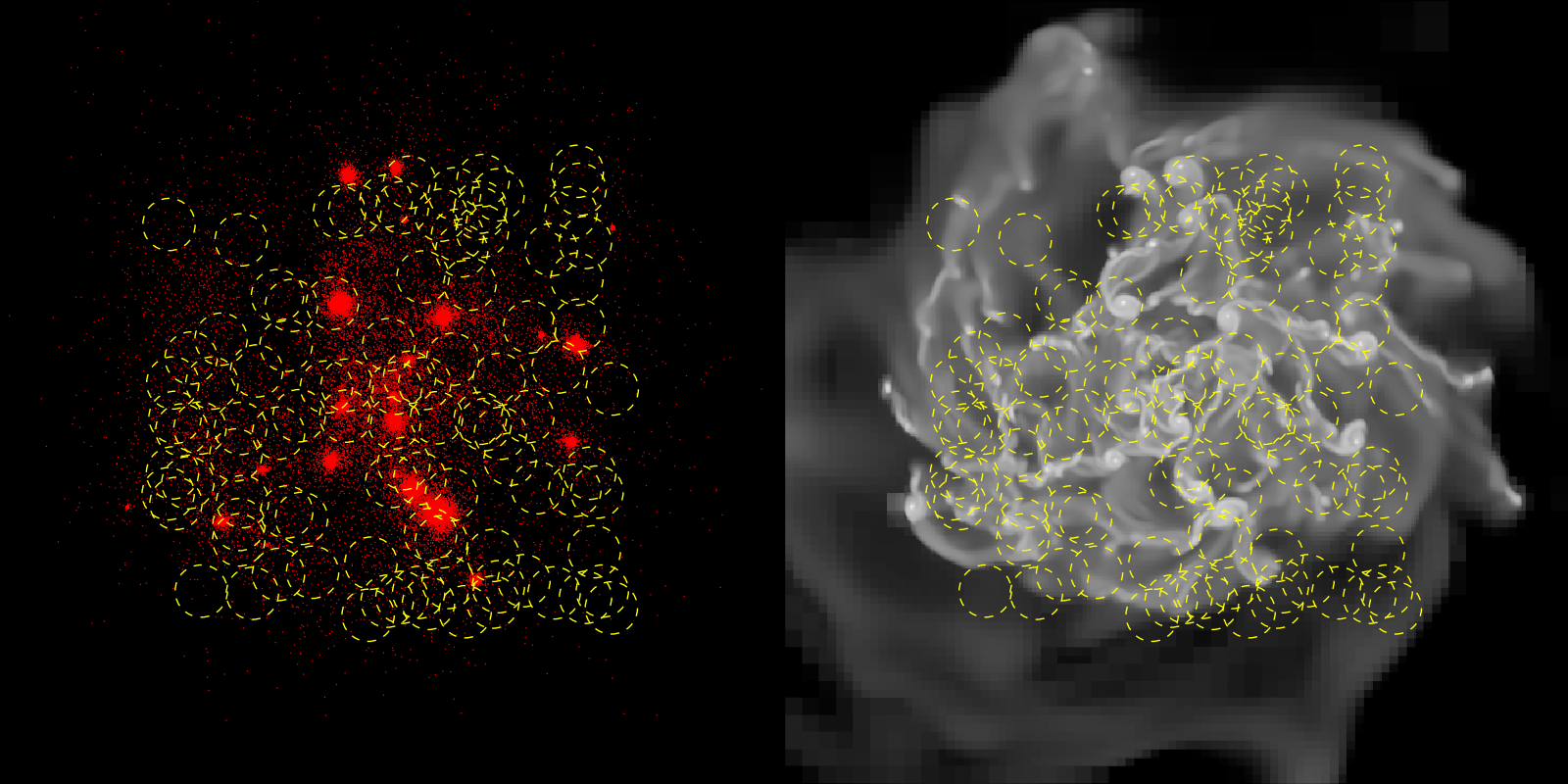}
	\caption{Face-on slice at the end of the run. Left: particles are plotted as red dots. Yellow dashed lines show the contours of the random disks with radius of 5 kpc used for this snapshot. Right: density slice at the same instant with the same contours overplotted. We can see that random disks cover a wide range of conditions, from low-density regions with just a few stars to stellar clumps located in high-density cells.}
	\label{fig:random}
	\end{center}
	\end{figure}

Motivated by these results, we explore the behavior of the KS relation at different scales. Previous studies have demonstrated that the KS relation does not hold at small surface densities \citep{Onodera_10, Verley_10}, so we perform a similar analysis in Figure \ref{fig:kssizes} to verify that result. In each panel we plot gas surface density and average SFR surface density for 100 points that represent disks at random locations in the galaxy disk. The radius of the disks are 100 pc, 500 pc, 1 kpc, 5 kpc, and 10 kpc for panels (a), (b), (c), (d), and (e) respectively. The vertical dotted line indicates $\sgas = 1~\msun \text{pc}^{-2}$, while the inclined dotted line is the original KS law. Colored dashed lines are the best fit for the high gas surface density range, with the values of their slopes showed at the lower right corner of each panel. An example of random disks is given in Figure \ref{fig:random} for the 1 kpc case (yellow dashed lines).

From the figure it is clear that, as the size of the regions increases, the slope of the best fit tends to a value close to 1.3. The most striking difference is obtained for the smallest size, 100 pc, in which case the slope is $n \sim 0.42$. We attribute this discrepancy to the limitations of our star formation algorithms. In Section \ref{subsec:SFandFeedback} we described a density threshold and a mass threshold for the creation of stars, motivated by previous observations in the case of the density, and to improve the simulation performance in the case of the mass. Since the minimum mass for a star particle is $10^5 \msun$, low SFRs due to low-mass star formation are suppressed at low gas surface densities. This translates into a saturation in the low densities regime, where most points lie in a roughly horizontal line. For greater sizes this feature disappears and we recover more reasonable values for the slope, converging to a value around 1.3 for sizes greater than a kiloparsec. At the same time, the scatter in the fit of each panel decreases as the size increases. Intuitively we know that for greater radii, the area of the galaxy disk enclosed by the disks will be similar for all points with a fixed radius. This is also the explanation of why in the case of panels (d) and (e) most of the points lie very close to each other. Based on this analysis, the slope $n = 1.16$ in Figure \ref{fig:kslaw} can be understood as the result of mixing different disk sizes, some of them with slopes considerable smaller than the \citet{K98} value.

Our results confirm observations of the KS relation at small regions, which say that this relation is not valid at those scales \citep{Onodera_10, Verley_10}. We can conclude then that the KS relation is valid for scales larger than $\sim$ 1 kpc, which can be interpreted as the characteristic scale of the problem. This result is concordance with the fact that surface densities are relevant only in scales much greater than the disk scale-height, which for our model is defined as 400 pc. Hence, it is safe to conclude that convergence is only achieved at scales greater than the scale-height. Therefore, for the rest of this study we will stick to a value $n \sim 1.3$, since it has been proven to be the convergent value of the slope for large disk sizes.

\subsection{Simulated Star Formation}
\label{subsec:simulatedsfr}

In the previous subsection we saw that similar values for both surface densities are obtained even when considering random regions of the disk with different sizes (Figure \ref{fig:kssizes}), which shows that a local relation might be predominant when averaging over greater regions. Inspired by those results, we have designed a numerical test to explore the plausibility of an intrinsic assumption in the star formation algorithm that allows the power law to be satisfied at all scales. This procedure allows us to study star formation using only the simulation outputs, with no need of the dynamical evolution of the system (and hence the ``simulated" name). In other words, when using this method we do not run the code again. We then proceed to calculate a simulated SFR based on the snapshots of the simulation, which are dumped every $\sim$4 Myr. For that purpose, we have set a few conditions that the cell must fulfill in order to create a star, similar to the method described in Section \ref{subsec:SFandFeedback}. We have tested constraints such as collapsing gas, cooling time, and Jeans mass, being the latter the one that most affects the resulting SFR. Since the Jeans criterion select cells with masses greater than a Jeans mass and $M_{\rm Jeans} \propto c^3_s \rho^{-1/2}$, the Jeans condition is equivalent to a density threshold for a given temperature. Our test therefore focuses on how the star formation law behaves if we assume only a density threshold and that the mass of the new star is given by a fraction of the mass of the cell $\mstar=\epsilon\rhogas\Delta x^3$, with no other physical criteria involved. The process to calculate the simulated SFR is as follows.

\begin{figure}[p]
	\begin{center}
	\includegraphics[scale=0.8]{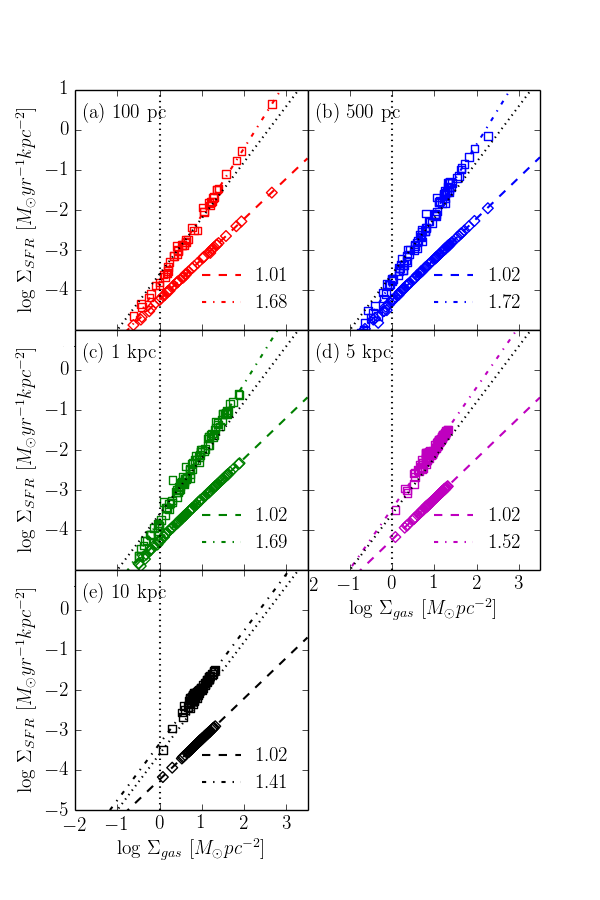}
	\caption{Influence of the timescale in the star formation law. Squares show SFRs computed using $\tdyn$, while SFRs calculated using a constant depletion time are plotted with diamonds. The best fit to the points is shown with dashed dotted and dotted lines respectively, with theirs slopes at the lower right corner of each panel. In all cases, the slope obtained using the dynamical time is greater, with values oscillating in the range 1.4 -1.7. On the other hand, when considering a constant $\tdep$ the slope get values around $n = 1$.}
	\label{fig:kssimulated}
	\end{center}
	\end{figure}

We select one of the last snapshots of the run. The calculation of the total mass of stars created is straightforward from the gas density and the size of the cells. A more complex computation is to decide which dynamical timescale is characteristic for star formation. A first method to calculate its value is to use the orbital time $\torb$, defined as the time needed by the gas to orbit at the galactic radius. Another approach is to consider the dynamical timescale as the free-fall time $\tff$, which represents the collapsing time of the gas. A third way is to assume that the gas is constantly depleted. In Figure \ref{fig:kssimulated} we present SFR surface densities for 100 random disks using the free-fall time (squares) and a constant depletion time of 500 Myr (diamonds). The panels show the same values for the radii of the disks as in Figure \ref{fig:kssizes}

The difference in slopes for both approaches is evident from the plot. The slope for the free-fall time varies between 1.4 and 1.7 depending on the size of the regions. Considering that $\tff \propto \rho^{-1/2}$ and $\rho_{\rm SFR} \propto \rho_{\rm gas} / \tff$, we can easily derived a relation of the type $\ssfr \propto \sgas^{1.5}$ for a fixed scale-height. We can say now that at large scales (5 and 10 kpc) the slope of the relation is close to the value just derived. On the other hand, we obtained a slope of $n \sim 1.02$ for the case of a constant depletion time. This exponent can also be understood from the fact that $\rho_{\rm SFR} \propto \rho_{\rm gas} / \tdep$, but as this time $\tdep$ is a constant, the dependency on the gas density remains linear. If a slope close to 1 is recovered when the star formation timescale does not depend on the density, then observations of such slopes suggest that star formation is gravitationally independent of the dynamics of gas. But, what processes produce the decoupling of star formation and gas dynamics? Many not-well-understood candidates appear, such as turbulence and magnetic fields, but a definite answer will require further investigation. In addition, our simulated star formation algorithm still depends on two parameters that may influence the slope, the efficiency $\epsilon$ and the density threshold $n_{\rm thres}$. The next step is then to study the impact of these parameters in the star formation laws.

\begin{figure}[h]
	\begin{center}
	\includegraphics[scale=0.6]{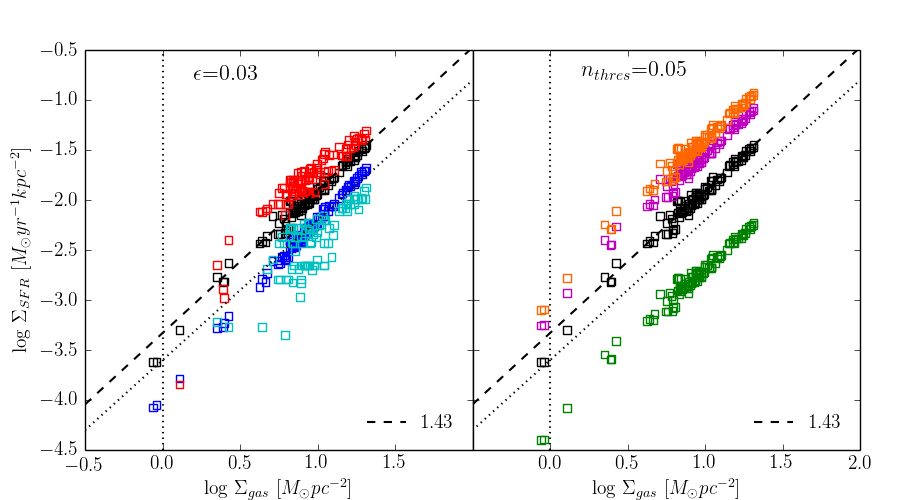}
	\caption{Dependence of the star formation laws on the efficiency and density threshold. On the left panel star formation efficiency is constant $\epsilon$ = 0.03 and the density threshold takes the values 0.0 (blue), 0.05 (black), 100 (red) and 1000 \cmmm~(cyan). Right panel show a constant threshold $n_{\rm thres}$ = 0.05 and efficiencies 0.005 (green), 0.03 (black), 0.07 (magenta) and 0.1 (orange). The dashed line show the best fit for the case $\epsilon$ = 0.03 and $n_{\rm thres}$ = 0.05. All other cases show a slope close to 1.43 differing between them in the zero point. For reference, the original Kennicutt-Schmidt fit is plotted in black dashed lines. A proper choice of both parameters allows us to fit the zero point of the star formation law.}
	\label{fig:simulatedneff}
	\end{center}
	\end{figure}

Figure \ref{fig:simulatedneff} shows the variation in the star formation law when changing the efficiency and the density threshold. We vary the density threshold, keeping the efficiency constant ($\epsilon$ = 0.03, left panel) and viceversa ($n_{\rm thres}$ = 0.05, right panel) for disks with a radius of 5 kpc. Results are shown for the SFR calculated using the dynamical time. For reference the original \citet{K98} fit is plotted in black dotted lines. For the constant efficiency case, we consider values of 0.0 (blue), 0.05 (black), 100 (red) and 1000 \cmmm~(cyan) for the density threshold. All different choices are consistent with a power-law with exponent $\sim 1.43$, in concordance with the KS value and close to the value of 1.3 deduced from Figure \ref{fig:kssizes}. This indicates that choosing the correct number for $n_{\rm thres}$ might account for the shift in the $y$-axis while keeping the slope constant. Regardless of the apparent success of this technique, a drawback can be concluded from the plot. Blue points indicate that the right slope is still recovered even when we consider a zero density threshold, so a threshold might not be as important as previously thought. Star formation efficiency then emerges as the main candidate to adjust the star formation law. We explore this possibility by keeping $n_{\rm thres}$ = 0.05 \cmmm~and using values of 0.005 (green), 0.03 (black), 0.07 (magenta), and 0.1 (orange) for the efficiency. Again the slope is in agreement with 1.43 and a change in efficiency might also account for the $y$-axis shift. One key aspect that demands our attention is the remarkable agreement in the slope between all cases. Picking out the correct pair of parameters always assures us consistency in the slope, implying that both variables are not independent and that a degeneracy relation exists between them.

We have discussed how the configuration of our experiment allows us to handle the efficiency as a mere numerical parameter to regulate the zero point of the star formation law. However, it is worth noting that the star formation efficiency is the result of the interaction of the many physical processes and scales involved in the formation of a star. From the assumption that star mass is a fraction of the gas mass, and considering a characteristic galactic timescale that correlates with gas surface density, we can easily derive a KS law that is dimensionally correct. This idea can be expressed in equations. Beginning with $\mstar=\epsilon\rhogas\Delta x^3$, we can derive the density of stars $\rhostar=\epsilon\rhogas$. Assuming a constant scale height, we get $\sstar=\epsilon\sgas$ and finally $\sstardot=\epsilon \sgas /\Delta \tau$, where $\Delta \tau = \Delta \tau (\sgas)$ is a galactic timescale that might be a function of the gas surface density . This implies that masses and timescales play a role in making the relation dimensionally consistent and that the physics of star formation is implicit in the efficiency. Our conclusions are still valid in that case, not as a parameter that we can manage but as the link between the ISM and star formation that determines the behavior of the star formation laws. It is thus the physical processes involved in the ISM-star formation relation, reflected in the efficiency, that regulate the SFR surface density-gas surface density relation. This issue have already been discussed in previous works \citep{Saitoh_08}. They obtained that a density threshold is the most important factor that determines the structure of galaxies. Although we see some differences in the case when $n_{\rm thres} = 0$, the plots show that there is no huge variation in the slope. So we are confident that a combination of both parameters is the key parameter to get the right star formation law.

\section{Summary and Conclusions}
\label{sec:conclusions}

Using the AMR code Enzo, we performed a simulation of an isolated local galaxy to study the ISM and star formation. The model includes diverse physical processes such as cooling and heating processes, star formation, and stellar feedback. The refinement criteria chosen allow us to reach a resolution of $\sim$ 40 pc, consistent with typical sizes of giant molecular clouds.

We let the disk evolve for $\sim 1$ Gyr, at which time we get a fragmented and highly turbulent disk with properties in agreement with a modern view of the ISM. A mix of cold high-density filaments and clumps coexist with hot low-density regions created by supernova explosions. Its density PDF is well fitted by a LN-PDF at high densities, with a characteristic density $\rhoo \approx 10^{-2.5}\msun \text{pc}^{-3}$ and a dispersion $\sigma_0 \approx 10^{1.0}\msun \text{pc}^{-3}$. At high masses the cell mass PDF is well fitted by a power law of exponent $\alpha = -1.39$, while the clump mass PDF shows a power-law consistent with $\alpha = -1.25$ at the high-mass end, which is slightly smaller than the value for the mass distribution of giant molecular clouds. The velocity power spectrum presents a double cascade, where energy is injected at scales of the order of $R_{\rm disk}$. At larger scales, there is an inverse energy cascade characterized by a Kolmogorov turbulence $E(k) \sim k^{-5/3}$, while at smaller scales an direct enstrophy cascade of the form $E(k) \sim k^{-3}$ is present. A better resolution is still necessary to fully characterize the ISM and to get a good description of turbulence and its relation to star formation at smaller scales.

Star formation in our simulation is well represented by a power law with a break at $\sgas = 1~\msun \text{pc}^{-2}$. At higher gas surface densities, the slope of the power-law is $n = 1.16$, while for lower values we get a greater exponent of $1.31$. When averaging over different disk sizes, we get that the slope tends to a value $\sim 1.3$ for scales greater than 1 kpc, suggesting this as the minimum scale at which the KS law is valid. Surprisingly, disks of all sizes lie in the same range in the $\ssfr - \sgas$ space, proposing a predominant local relation that also holds when averaged globally.

Motivated by the previous result, we investigate if our assumptions in the star formation algorithm at small scales dominate over global dynamics. For this purpose, we construct a methodology to calculate SFRs with no other criteria than those restrictions. In such a case we do not need to rerun the code since our method is devised to work only with simulation outputs. In our algorithm we define that the mass of the star formed is equal to the mass of the cell multiplied by the star formation efficiency, so the calculation of the total star mass is straightforward. But a timescale is still needed to calculate the SFR. We chose to make a comparison between using the free-fall time $\tff$ and a constant depletion time $\tdep$ as star formation timescales. The former gives higher slopes, converging to $\sim$ 1.4 for high surface densities, while the latter gives slopes close to 1 in all cases. This suggests that a decoupling between star formation and gas evolution might be the cause for a lower slope.

Another important fact to consider is the criteria that a cell has to satisfy to create a star in our simulated star formation. We find that convergent gas and cooling time have no major relevance in shaping the star formation law. On the other hand, changing the Jeans mass and the density threshold criterion can produce an increase/decrease in the amount of stars formed. Since both are related by $M_{\rm Jeans} \propto \rho^{-1/2}$, we pick the density threshold as a representative parameter for both conditions. In addition, the efficiency used to transform gas mass to star mass remains as a free parameter that could change the results. We explore the variations in the star formation law by keeping one of those parameters constant, while changing the other. We get a consistent slope in all cases independently of the values of $\epsilon$ and $n_{\rm thres}$. We then conclude that the assumption that stars are created with a fraction of the mass of the parent cell is enough to recover the star formation law. No other physical criteria, such as collapsing gas, cool gas, or Jeans instability are needed to agree with a power law. This can be easily derived analytically from the expression $\mstar=\epsilon\rhogas\Delta x^3$ assuming a constant scale-height and a galactic timescale $\Delta \tau$. A power-law with a slope similar to a KS law is thus a natural outcome of assuming that mass of stars is proportional to the mass of the cell. These results are remarkable considering that in this method we lose considerable information between snapshots. Moreover, we get these conclusions using a single snapshot, with no further information needed. From our test we then deduce that star formation efficiency is an essential parameter to determine the normalization of the star formation law. However, efficiency is not just a free parameter to fit the curve. We can go back to the definition of star formation efficiency and consider it as the outcome of the interplay between the physical processes governing star formation. In such a case those processes are implicitly present in the star formation algorithm and hence they ultimately regulate the star formation process.

Our conclusions give insights of why previous works have recovered a star formation law despite their differences in the treatment of other processes. They all use this numerical assumption, either explicitly \citep{Kravtsov_03, Tasker_Bryan_06, Robertson_08} or implicitly in a sink particle formulation \citep{Li_05}. Our interpretation also explains recent numerical works that have explored the dependence of SFR on physical constraints. \citet{Hopkins_13} varied the star formation restrictions in a set of runs using the same code in all of them. They found that criteria such as self-gravity, density threshold, molecular gas, temperature, Jeans instability, converging flows, and rapid cooling do not significantly change the SFR, in agreement with our analysis. So far, we have considered that gas is consumed only to form stars. A particularly interesting situation to study is then the presence of a new source of gas consumption (e.g. the central massive black holes). If that is the case and an important fraction of the gas is accreted into the new sink particle, then that accreted mass also needs to be considered to satisfy a star formation law of the type $\ssfr - \sgas /\torb$ \citep{Escala_07}.

A key question now arises: which process is most important in regulating star formation efficiency? Observations have not successfully addressed this question due to either resolution limits or because observational errors from the multiple processes that tends to cancel each other out and cannot be decoupled. It is then necessary to observe extreme star formation environments where the efficiency suffers a significant change, for instance in high-redshift starburst. In such a case, the bi-modality found by \citet{Daddi_10, Genzel_10} might be explained by a low-/high-efficiency double regime. Moreover, it must be considered whether second parameters may affect the efficiency and hence shape the KS law. So far, we have assumed a global value for the efficiency, but many previous studies have attempted to find this value \citep{Krumholz_Tan_07, Hennebelle_11, Padoan_11} with no general consensus. A straightforward assumption is therefore that efficiency should depend on some local properties \citep[e.g.,][]{Federrath_12}, for instance cloud-cloud collisions \citep{Tan_00}, molecular gas fraction \citep{Krumholz_McKee_Tumlinson_09}, or collapsing molecular clouds \citep{Zamora-Aviles_12}. Extending this assumption leads us to question whether the existence of second parameters at larger scales might also be possible. A particularly appealing galactic quantity for second parameters is the largest scale not stabilized by rotation. This is the only well-defined galactic-scale value in the gravitational instability problem \citep{Escala_08}. It also correlates with global SFR \citep{Escala_11} and it can even be generalized for the case without large-scale rotation \citep[e.g., in galaxy merger;][]{Escala_13}. Despite our findings, further investigation is still needed in order to come up with a full description of star formation and its relation with the ISM.

We would like to thank the anonymous referee for helpful suggestions and comments that improved the paper. F.B. acknowledges support from Programa Nacional de Becas de Postgrado (grant D-22100632). A.E. acknowledges partial support from the Center for Astrophysics and Associated Technologies CATA (PFB 06), Anillo de Ciencia y Tecnolog\'ia (Project ACT1101), and Proyecto Regular Fondecyt (grant 1130458). The simulation was performed using the HPC clusters Markarian (FONDECYT 11090216) and Levque (ECM-02). The analysis and plots were carried out with the publicly available tool $\mathtt{yt}$ \citep{Turk_11}.

\bibliographystyle{apj}
\bibliography{./bibliography}

\end{document}